\documentclass[12pt]{iopart}
\usepackage{epsfig}
\usepackage{amssymb}
\usepackage{mathrsfs}
\def\shiftleft#1{#1\llap{#1\hskip 0.04em}}
\def\shiftdown#1{#1\llap{\lower.04ex\hbox{#1}}}
\def\thick#1{\shiftdown{\shiftleft{#1}}}
\def\b#1{\thick{\hbox{$#1$}}}
\begin{document}

\title[Coulomb and CSB effects in the three-nucleon system]
{ The properties of the three-nucleon system with the dressed-bag model for 
 $NN$ interaction. II: Coulomb and CSB effects}
\author{V. I. Kukulin, V. N. Pomerantsev}
\address {Institute of Nuclear Physics, Moscow State University, 119899 Moscow, Russia}
\author{and Amand Faessler }
\address {Institut f\"ur Theoretische Physik, Universit\"at T\"ubingen\\ 
Auf der Morgenstelle 14, D-72076 T\"{u}bingen, Germany}
%\maketitle
\begin{abstract}

Coulomb and charge-symmetry breaking effects in the $^3$He ground state within
the dressed dibaryon model developed recently for  $2N$ and $3N$ forces are
examined in detail. Particular attention has been paid to the Coulomb 
displacement energy $\Delta E_C=E_B(^3{\rm H})- E_B(^3{\rm He})$ and rms charge
radii of $^3$H and  $^3$He. A new scalar $3N$ force between the third nucleon and
dibaryon is found to be very important for  a correct description of the Coulomb
energy and rms charge radius in $^3$He. In view of the new  results for $\Delta
E_C$ obtained here, the role of the effects of charge symmetry breaking in the 
nuclear force is discussed.

\end{abstract}

%\begin{keyword}

%\PACS
%\end{keyword}

\section{Introduction}

The problem of accurate description of Coulomb effects in $^3$He in the
current $3N$ approach of the  Faddeev or variational type has attracted much
attention for last three decades (see e.g.~\cite{Friar,Gloeckle}  and the
references therein to the earlier works). The $\Delta E_C$ problem dates back
to the first accurate $3N$  calculations performed on the basis of the Faddeev
equations in the mid-1960s~\cite{Gignoux}. These  calculations first exhibited
a hardly removable difference of ca. 120 keV between the theoretical
prediction  for $\Delta E_C^{\rm th} \simeq 640$~keV and the respective
experimental value $\Delta E_C^{\rm exp}  \simeq 760$~keV. In subsequent 35
years, numerous accurate $3N$ calculations were performed over the  world
using many approaches, but this puzzle was still generally unsolved. The most
plausible quantitative explanation (but yet not free of serious questions) for
the puzzle has  been recently suggested by Nogga et al.~\cite{Gloeckle}. They
have observed that the difference in  the singlet $^1S_0$ scattering lengths
of $pp$ (nuclear part) and $nn$ systems (originating from the  effects of
charge symmetry breaking (CSB)) can increase the energy difference between
$^3$H and $^3$He  binding energies and thus contribute to $\Delta E_C$. Using
some realistic, currently accepted,  $a_{pp}$ and $a_{nn}$ values and other
small corrections, Nogga et al.~\cite{Gloeckle} were able to  virtually remove
the gap of 120~keV between the conventional $3N$ approaches (which neglect
CSB  effects) and the experimental value. However, this success depends
crucially on the accepted $a_{nn}$  value, which is not very reliable up to
date (see the details in Section IV). For another admissible  $a_{nn}$ value,
the explanation given for the gap in ref.\cite{Gloeckle} is invalid. Thus, one
should  look for another, alternative, explanation for the puzzle.

In this paper, we give such an alternative explanation of the $\Delta E_C$ puzzle and
other Coulomb  effects in $^3$He without any free parameter on the basis of the new
model recently developed for the  $2N$ and $3N$ forces by our joint Moscow-T\"ubingen
group~\cite{IJMP,Jphys02}. The model includes an  inevitable dibaryon in the
intermediate state dressed with $\pi$, $\sigma$, $\omega$, and $\rho$  fields,
together with the traditional Yukawa $\pi$ and $2\pi$ exchanges, which describe the
peripheral  part of the $NN$ interaction. Being embedded into three- and many-body
systems, this specific two-body  mechanism generates an inevitable scalar three-body
force induced by $\sigma$-meson exchange between  the dressed dibaryon and
surrounding nucleons~\cite{Ku03}. 

In the preceding paper~\cite{Ku03}, we formulated a new model for the $3N$
force of the scalar nature  and tested it in $3N$ calculations. As was
demonstrated in~\cite{Ku03}, this $3N$ force is  so strong that can explain not
only the $3N$ binding energy but also other important characteristics  of the
$3N$ system. This scalar $3N$ force is closely associated with the generation 
of an intermediate dressed dibaryon in the fundamental $NN$
interaction~\cite{IJMP,Jphys02}. The contribution of the  above scalar $3N$
force between the  dressed dibaryon and third nucleon to the total $3N$ binding
energy is much higher than that of the  conventional $3N$ force associated with
the generation of intermediate $\Delta$-isobars and two-pion  exchanges between
three nucleons (see e.g. the review paper~\cite{3NF}  and the references therein
to earlier works). Thus, this scalar $3N$ force should primarily determinate
the  properties of the $3N$ nuclei, e.g., the rms radii of matter and charge
distributions, the probability  of the $D$ state, the constants of the
asymptotic normalization in the $S$- and $D$-wave components,  etc. We also
established that the complicated interplay between $2N$ and $3N$ forces in the
new model is primarily responsible for Coulomb effects in $^3$He and the
Coulomb  displacement energy $\Delta E_C$.

This paper is devoted to the Coulomb and CSB effects in $^3$He, which provide
an independent and important  check for the consistency and adequacy of the
new force model. We found that all basic Coulomb effects  can be quite
naturally explained in the framework of the new force model without any
additional  parameter. Thus, this independent check provides quite strong
additional support for the force model  used here.

The paper is organized as follows. In Sect.~II, we present some theoretical
framework for treating  Coulomb effects within the new force model using the
isospin formalism. Section~III is devoted to  calculation results for the
$^3$He ground state. The detailed discussion of the results obtained is
presented in Sect.~IV,  while Sect.~V incorporates the concluding remarks of
the study.

\section{Theoretical framework}

In this section, we present a necessary formalism for our variational
calculations of $3N$ systems  within the framework of the multicomponent
dressed-bag model (DBM). In addition, we discuss here the  details concerning
the inclusion of Coulomb effects and determination of observables in the
$6qN$  channel. The DBM described in detail in~\cite{IJMP,Jphys02} differs from the
traditional OBE-type $NN$  interaction models primarily by the presence of
{\em non-nucleonic components} in the nuclear wave function.  Contrary to
numerous hybrid quark-nucleon models (popular in the 1980s), it {\em
explicitly} involves  mesonic degrees of freedom inside the dressed bags.
Hence, we consider the $3N$ system, where the  traditional $3N$ channel is
supplemented by three other channels involving the dressed dibaryon  (dressed
six-quark bag) interacting with the third nucleon. In the case of $^3$He, a
large scalar  force appears due to $\sigma$-meson exchanges between the
dibaryon and extra nucleon, and the  additional Coulomb force arises because
the bag and rest nucleon can have an electric charge. This  {\em new Coulomb
three-body force} is responsible for a significant part of the total $^3$He
Coulomb  energy (this three-body Coulomb force has been missed fully in
previous $3N$ calculations within hybrid $6qN$ models~\cite{Dijk}). It should
be emphasized here that the contribution of this three-body Coulomb force to
the  total three-body binding energy is (as will be demonstrated below) quite
significant ($\sim 100$~keV)  and makes it possible to explain, in essence,
the experimental $\Delta E_C$ value.

The second feature of the interaction model used here is the absence of the
local $NN$ short-range  repulsive core. The role of this core is played by the
condition of orthogonality to the confined $6q$  states forbidden in the $NN$
channel. (These states can be identified, e.g., with locked colour $6q$ 
states having the tetraquark-diquark structure~\cite{tetra}.) This
orthogonality requirement imposed on the relative-motion $NN$ wavefunction is
responsible for the  appearance of some inner nodes in this wavefunction (very
stable under variation of the $NN$-channel  energy\footnote {The fact that
these stationary nodes play role of the repulsive core in traditional 
$NN$-force models has long been established~\cite{nodes}}) and respective
short-range loops. These  short-range nodes and loops lead to numerous effects
and general consequences for the nuclear  structure (see below). One of these
consequences is a rather strong overestimation of the Coulomb contribution
when using the Coulomb interaction between  point-like nucleons. Thus, it is
necessary to take into account the finite radius of the nucleon  charge
distribution.

\subsection{Construction of a $3N$ variational basis and the wavefunction of the
$6qN$ component}

Here, we give the form of the basis functions used in this work and the corresponding
notation for the  quantum numbers. The total wavefunction of the $3N$ channel,
$\Psi_{3N}$, can be written in the  antisymmetrized basis as a sum of the three
components:
\begin{equation}
 \Psi_{3N}=  \Psi^{(1)}_{3N}+ \Psi^{(2)}_{3N} + \Psi^{(3)}_{3N},
 \label{psi3n}
 \end{equation}
where the label (i) enumerates one (of three) possible set of the Jacobi coordinates
$({\bf r}_i,{\b  \rho}_i)$. Every component in eq.(\ref{psi3n}) takes the form 
\begin{equation}
  \Psi^{(i)}_{3N}=\sum_{\gamma}\sum_{n=1}^{N_{\gamma}}
  C_{\gamma n} \Phi^{(i)}_{\gamma n}.
 \label{psi3ni}
 \end{equation}
The basis functions $\Phi^{(i)}_{\gamma n}$ are constructed from Gaussian functions
and the  corresponding spin-angular and isospin factors:
\begin{equation}
 \Phi^{(i)}_{\gamma n}=\exp\{-\alpha_{\gamma n}r_i^2 -
 \beta_{\gamma n}\rho_i^2\} {\cal
 F}_{\gamma^{(i)}}(\hat{r}_i,\hat{\rho}_i,\{\xi\}_i){\cal T}_{\gamma^{(i)}},
 \label{Phigamn}
 \end{equation}
 where the composite label $\gamma^{(i)} =\{\lambda_i\,l_i\,L\,S_{jk}\,S\,t_{jk}\}$
represents the respective set of the quantum numbers for the basis functions
(\ref{Phigamn}):  
$\lambda_i$ is the orbital angular momentum of the $jk$ pair; $l_i$
is the orbital angular momentum of  the third nucleon ($i$) relative to the center of
mass of the $jk$ pair; $L$ is the total orbital  angular momentum of the $3N$ system;
$S_{jk}$ and $t_{jk}$ are the spin and isospin of the $jk$ pair,  respectively; and
$S$ is the total spin of the system. We omit here the total angular momentum $J=1/2$ 
and its $z$-projection $M$, as well as the total isospin of the system $T=1/2$ and
its projection  $T_z$ (in this work, we neglect the very small contribution of the
$T=3/2$ component). The spin-angular and isospin parts of the basis functions are
taken in the form
\begin{equation}
 {\cal F}_{\gamma^{(i)}}=|\{\lambda_i l_i:L\}\{s_js_k(S_{jk})s_i:S\}:JM\rangle
 \label {fgi}
 \end{equation}

 \begin{equation}
 {\cal T}_{\gamma^{(i)}}\equiv {\cal T}_{t_{jk}} = |t_jt_k(t_{jk})t_i:TT_z\rangle
 \label {tgi}
 \end{equation}
 
The nonlinear parameters of the basis functions $\alpha_{\gamma n}$ and
$\beta_{\gamma n}$ are chosen  on the Chebyshev grid  which provides the completeness
of the basis and fast convergence of variational calculations~\cite{Cheb}. As was
demonstrated earlier~\cite{Gauss}, this few-body Gaussian basis is  very flexible and
can represent rather complicated few-body correlations. Therefore, it leads to quite
accurate  eigenvalues and eigenfunctions. The formulas for matrix elements of
Hamiltonian (for local $NN$ interactions) with antisymmetrized Gaussian basis
are given in paper~\cite{Tur1}.

Having the three-nucleon component $\Psi_{3N}$ found in the variational calculation,
one can construct  the $6qN$-channel wavefunction $\Psi_{6qN}^{(i)}$, which depends on
the coordinate (or momentum) of the  third nucleon and the $\sigma$-meson momentum
and includes the bag wavefunction (see eq.(33) of  ref.\cite{Ku03}). Integrating the
modulus squared of this function with respect to the meson momentum  and inner
variables of the bag, one obtains the density distribution of the third nucleon
relative to  the bag in the $6qN$ channel. This density can be used to calculate all
observables whose operators  depend on the variables of the nucleons and bag.
However, it is much more convenient and easier to  deal with the quasi-wavefunction
of the third nucleon in the $6qN$ channel, which has been introduced  by eq.(39) of
ref.~\cite{Ku03}.

To calculate matrix elements of 3BF Coulomb and OPE forces one needs the spin-isospin
part of $6qN$ components of the total wavefunction. Here we give them
explicitly. The potential form factors in nucleon channels
$\varphi^{J_iM_i}_{L_i}$ now include the spin-isospin part with quantum numbers
of the dressed bag:
 \begin{equation}
\varphi^{J_iM_i}_{L_i}= \phi_{J_iL_i}({p}_i) {\cal
Y}^{J_iM_i}_{L_iS_d}(\hat{p}_i){\cal T}_{t_d},
 \label {phijl}
 \end{equation}
where  $S_d$ and $t_d$ denote the  bag spin and isospin respectively, and $J_i$
and $L_i$ are the  total and orbital angular momenta, respectively, referring
just to the vertex form factor (the  additional letter ``i'' is introduced to
distinguish these quantum numbers from the respective total  angular momentum
$J$ and orbital angular momentum $L$ of the whole system). Since the present
version of the DBM involves bag states with zero orbital angular momentum
(although  a more general treatment can also include dressed-bag components
with $L_d\ne 0$), we have $J_i=S_d$,  while the bag spin and isospin are
opposite to each other: $t_d+S_d=1$. The isospin part of the form factor is
\[ {\cal T}_{t_d}=|{\case{1}{2}}{\case{1}{2}}:t_dt_{d_z}\rangle .\] 

The total overlap function (with its spin-isospin part) 
 \begin{equation}
 \chi^{J_iM_i}_{L_i}(i)=\langle\varphi^{J_iM_i}_{L_i}|\Psi_{3N}\rangle
 \label {chijl}
 \end{equation}
can be written, e.g., as (cf. eq.(35) in ref.~\cite{Ku03})
\begin{equation}
 \fl \chi^{J_iM_i}_{L_i}({\bf q}_i)=
 \sum_{\lambda_i}\Phi^{J_iL_i}_{\lambda_i{\cal J}}(q_i)
 \langle {\cal J}m_{\cal J}J_iM_i|JM\rangle \,
 {\cal Y}^{{\cal J}m_{\cal J}}_{\lambda_i {\case{1}{2}}}(\hat{\bf q}_i)
 \,\langle t_dt_{d_z}{\case{1}{2}}t_{z_i}|TT_z\rangle
 \,{\cal T}_{{\case{1}{2}}t_{z_i}}.
 \label {ovfull}
 \end{equation}
 Here, $J$ and $M$ are the total angular momentum of the $3N$ system and its
projection, $\lambda_i$ and $\cal J$ are  the orbital and total angular 
momenta of the third ($i$th) nucleon respectively, ${\cal
T}_{{\case{1}{2}}t_{z_i}}$ is isospinor corresponding to the  third
nucleon. In the present calculations of the ground states of $^3$H and $^3$He 
($J=1/2$), we considered only the two lowest even partial waves ($S$ and $D$)
in $3N$  wavefunctions.  Therefore,  $\lambda_i$ can take only two values 0 or
2. Moreover, the total angular momentum of third nucleon $\cal J$ is uniquely
determined by value of $\lambda_i$: ${\cal J}=1/2$ at $\lambda_i=0$ and ${\cal
J}=3/2$ at  $\lambda_i=2$. So, we did not include the sum over $\cal J$ in
eq.(\ref{ovfull}).

Now we redefine the quasi-wavefunction of the third nucleon in $6qN$ channel,
including in it spin-isospin part of the bag:
\begin{equation}
 \tilde{\Psi}^{6qN}_{J_iL_i}=\sum_{M_it_{d_z}} 
 \sqrt{-\frac{\rm d}{{\rm d}E}\lambda^{J_i}_{L_iL_i}(E-\frac{q_i^2}{2m})}
 \chi^{J_iM_i}_{L_i}\,|J_iM_i\rangle \, {\cal T}_{t_dt_{d_z}},
 \label {psi6qn}
 \end{equation}
where $|J_iM_i\rangle$ and  ${\cal T}_{t_dt_{d_z}}$ are spin and isospin parts of
the bag function respectively.

  It is easy to see that the three form factors
$\varphi^{J_i}_{L_i}$ used in this work  ($\varphi^{0}_0$, $\varphi^1_0$, and
$\varphi^1_2$) determine five radial components of the overlap  function
$\Phi^{J_iL_i}_{\lambda_i{\cal J}}(q_i)$ and five respective components of the
quasi-wavefunction  for the $6qN$ channel. To specify these components it is
sufficient to give three quantum numbers, e.g. $S_d$, $\lambda_i$ and $L_i$, and
 we will use notation 
 $\Psi^{6qN}_{S_d,\lambda_i,L_i}(q_i)$ for these radial component:

\[
\begin{array}{lcl}
 \Psi^{6qN}_{00,0}& :&(J_i=S_d=0,\,t_d=1,\,L_i=0,\,\lambda_i=0,\,{\cal J}=\frac{1}{2})\\
 \Psi^{6qN}_{10,0}& :&(J_i=S_d=1,\,t_d=0,\,L_i=0,\,\lambda_i=0,\,{\cal J}=\frac{1}{2})\\
 \Psi^{6qN}_{12,0}& :&(J_i=S_d=1,\,t_d=0,\,L_i=0,\,\lambda_i=2,\,{\cal J}=\frac{3}{2})\\
 \Psi^{6qN}_{10,2}& :&(J_i=S_d=1,\,t_d=0,\,L_i=2,\,\lambda_i=0,\,{\cal J}=\frac{1}{2})\\
 \Psi^{6qN}_{12,2}& :&(J_i=S_d=1,\,t_d=0,\,L_i=2,\,\lambda_i=2,\,{\cal J}=\frac{3}{2})
 \end{array}
\]
At last, we give the formula for the total quasi-wavefunction of $6qN(i)$ 
component, separating
explicitly its spin-angular and isospin parts:
\begin{equation}
 \Psi_{6qN}^{(i)}=\sum_{\lambda_iS_d}\left
 \{\sum_{L_i} \Psi^{6qN}_{S_d,\lambda_i,L_i}(q_i)\right \} 
 |\lambda_i {\case{1}{2}}({\cal J})S_d:JM \rangle \, 
 |t_d{\case{1}{2}}:TT_z\rangle.
 \label {psi6qntot}
 \end{equation}
The explicit dependence of this function on the isospin projection $T_z$ is
important for calculation of Coulomb matrix elements and rms charge radius. 

The interaction matrix elements include the overlap integrals of the potential
form factors with the  basis functions $\Phi_{\gamma ,n}= \Phi^{(1)}_{\gamma
,n} + \Phi^{(2)}_{\gamma ,n} + \Phi^{(3)}_{\gamma ,n}$, where all five above 
components of the overlap function enter into the matrix elements independently
(certainly, some of  the matrix elements can vanish). The explicit formulas for
the above overlap functions and detailed formulas for the  matrix elements of
all DBM interactions will be published elsewhere. However, when calculating
both the normalization of the $6qN$ component and observables, the $6qN$ 
components distinguishing only by their radial parts, i.e. by only $L_i$ ,
can be summed. Thus, only
three components of the  $6qN$ wavefunction (orthogonal due to their
spin-angular parts) remain: $S$-wave singlet one
$(S_d=0)$:
\[\Psi_{00}^{6qN} \equiv \Psi^{6qN}_{00,0} \]
and two triplet $(S_d=1)$ ones:
\[\Psi_{10}^{6qN} = \Psi^{6qN}_{10,0} +  \Psi^{6qN}_{10,2}, \]
\begin{equation} \Psi_{12}^{6qN} = \Psi^{6qN}_{12,0} +
\Psi^{6qN}_{12,2}.
 \label{3comp}
\end{equation}
The total weight of each of three $6qN(i)$,(i=1,2,3) components is equal to
\begin{equation} P_{6qN}^{(i)} = \|\Psi^{6qN}_{00}\|^2 +
\|\Psi^{6qN}_{10}\|^2 + \|\Psi^{6qN}_{12}\|^2; \qquad i=1,2,3.
 \label{P6qNi}
\end{equation}
Now, let us introduce the relative weights of individual $6qN$ components:
\begin{equation}
 P^{S0}_{6qN} =\frac {\|\Psi^{6qN}_{00}\|^2}{P^{(i)}_{6qN}}, \;
 P^{S1}_{6qN} =\frac {\|\Psi^{6qN}_{10}\|^2}{P^{(i)}_{6qN}}, \;
 P^{D}_{6qN} = \frac {\|\Psi^{6qN}_{12}\|^2}{P^{(i)}_{6qN}}, \;
 \label{P6qNrel}
\end{equation}
After renormalization of the total four-component wavefunction, the total weight of all $6qN$ 
components is equal to
\begin{equation}
 P_{6qN} =\frac {3P^{(i)}_{6qN}}{1+3P^{(i)}_{6qN}}
 \label{P6qN}
\end{equation}
Here we assume that the $3N$ component of the total wavefunction, 
$\Psi^{3N}$, obtained from the variational calculation is normalized to unity
while the total weight of the three-nucleon component $\Psi_{3N}$ is equal to
\begin{equation}
 P_{3N} =\frac {1}{1+3P^{(i)}_{6qN}}=1-P_{6qN}.
 \label{P3N}
\end{equation}

The total weight of the $D$-wave component in $^3He$ (and $^3H$ wavefunctions with allowance 
for non-nucleonic components is also changed:
\begin{equation}
 P_{D} = P_{3N}^{D}(1-P_{6qN}) + P_{6qN}^{D}P_{6qN}
 \label{PD}
\end{equation}

Numerical values of all above probabilities for  $6qN$ and $3N$ components are given below in
Table~2.  The total weight of all $6qN$ components $P_{6qN}$ in the $3N$ system, as was
demonstrated  in~\cite{Ku03}, is rather large and approaches or even exceeds 10\%. Furthermore,
taking into account  the short-range character of these components, the more hard nucleon momentum
distribution (closely  associated with the first property) for these components, and very strong
scalar three-body  interaction in the $6qN$ channel, one can conclude that these non-nucleonic
components are very  important for the properties of nuclear systems .

\subsection{``Smeared'' Coulomb interaction}
The Gaussian charge distribution that has the rms charge radius $r_c$ and is normalized to the total 
charge $z$: $ 4\pi\int\rho r^2dr = z$  can be written as
\begin{equation}
 \rho(r)=z\left (\frac{\alpha}{\pi}\right )^{3/2}{\rm e}^{-\alpha r^2}, 
 \qquad \alpha^{-1}=\frac{2}{3}r_c^2.
 \label{rhogauss}
\end{equation}
The Coulomb potential for the interaction between such a charge distribution $\rho(r)$ and a point-
like charged particle has the well-known form
\[
 V(R;\alpha)= \int\frac{d{\bf r}\, \rho(r)}{|{\bf R-r}|} =
 \frac{z}{R}{\rm erf}(R\sqrt{\alpha})
\]
One can derive a similar formula for the Coulomb interaction between {\em two} charges $Z_1$ and $Z_2$ with 
Gaussian
distributions  with different widths $\alpha_1$ and $\alpha_2$ and rms radii $r_c^1$ and $r_c^2$,
respectively:
 \begin{equation}
 V(R;\alpha_1,\alpha_2) = \frac{z_1z_2}{R}{\rm erf}(R\sqrt{\tilde{\alpha}}),
 \, \tilde{\alpha} =\frac{\alpha_1\alpha_2}{\alpha_1 + \alpha_2}, \mbox{ or }
 \tilde{\alpha}^{-1}=\frac{2}{3}(r_{c_1}^2 + r_{c_2}^2).
 \label{Vcoul}
\end{equation}
In our calculations, we used the following charge radii for the nucleon and dibaryon:
\[(r_c)_p =0.87\mbox { fm},\]
\[(r_c)_{6q}=0.6\mbox{ fm}.\footnotemark{} \]
\footnotetext{%
This value is simply the rms charge radius of the six-quark bag with the
parameters given in  ref.~\cite{IJMP}. The neutral $\sigma$ field of the bag
changes this value only slightly. The evident  difference between the charge
radii of the nucleon and dibaryon can be well understood as follows: the 
charge radius of the $3q$ core of the nucleon is taken usually as
$r_c^{3q}\simeq 0.5 \div 0.55$~fm,  while remaining 0.3~fm is assumed to come
from the charge distribution of the $\pi^+$ cloud  surrounding the $3q$ core in
the proton. In contrast, the meson cloud of the dibaryon in our approach  is
due to the neutral scalar-isoscalar $\sigma$ meson, so that the dibaryon charge
distribution is  characterized only by the charge radius of the bare $6q$
core. However in more complete  model, one should incorporate also $\pi$-meson
(and also $\rho$- and $\omega$-meson) dressing of the dibaryon. So that, the charge
radius of the ($pn$) and ($pp$) dibaryons will be a bit larger than the value 
0.6~fm  accepted here}.

These values lead to the ``smeared'' Coulomb interactions of the form
\[
 V^{\rm Coul}_{NN}(r) = \frac{e^2}{r}{\rm erf}(r\sqrt{\alpha_{NN}}), \qquad 
 \alpha_{NN}^{-1/2}=1.005\mbox{ fm};
 \]
  \begin{equation}
  V^{\rm Coul}_{6qN}(\rho) = \frac{e^2}{\rho}{\rm erf}(\rho\sqrt{\alpha_{6qN}}), 
\qquad \alpha_{6qN}^{-1/2}=0.863\mbox{ fm};
 \label{Vcoul1}
\end{equation}

\subsection{Matrix elements of the three-body Coulomb force}

The Coulomb interaction between the charged bag and third nucleon in the $3N$ channel is determined
by  the three-particle operator with the separable kernel (see eq.(40) in ref.\cite{Ku03}):
\begin{eqnarray}
 \fl ^{\rm 3BF}_{\rm Coul}V^{(i)}({\bf p}_i,{\bf p'}_i;{\bf q}_i,{\bf
 q'}_i)\nonumber \\
 \fl  = \sum_{J_iM_i,L_iL'_i}
  \varphi^{J_iM_i}_{L_i}({\bf p}_i)\,
  {}^{\rm Coul}W^{J_i}_{L_iL'_i}({\bf q}_i,{\bf q'}_i;E)\,
  \varphi^{J_iM_i}_{L'_i}({\bf p'}_i)\,
  \frac{1\!+\!\tau_3^{(i)}}{2}\frac{2\!+\!\tau_3^{(j)}\!+\!\tau_3^{(k)}}{2},
 \label{V3bfCoul}
\end{eqnarray}
where ${\bf p}_i$ is the relative momentum in the $jk$ pair, ${\bf q}_i$ is the
third nucleon  momentum, $E$ is the total energy of the $3N$ system. The
kernel  function $^{\rm Coul}W^{J_i}_{L_iL'_i}$ for the point-like Coulomb
interaction can be taken for the OSE interaction from eq.(53) of ref.[1] with
the substitution  $m_{\sigma}\Rightarrow 0$ and $-g^2_{\sigma} NN \Rightarrow
e^2$. Variational calculations require only the matrix elements (m.e.) of the
interaction operator between the basis functions chosen. It is evident that the
m.e. of the operator  (\ref{V3bfCoul}) can be expressed in terms of the
integrals of the overlap functions $\chi^{J_iM_i}_{L_i}({\bf q}_i)$ (\ref{ovfull}): 
\begin{equation}
\fl  \langle {\rm Coul\,3BF}\rangle_p = e^2\sum_{J_iL_iL'_i}
\lambda^{J_i}_{L_iL'_i}(0)
  (1+a)\frac{1}{1+t_d}
  \int \frac{\chi^{J_iL_i}({\bf q})}{E\!-\!E_0\!-\!\frac{q^2}{2m}}
 \, \frac{1}{({\bf q}\!-\!{\bf q'})^2}\, \frac{\chi^{J_iL'_i}({\bf
  q'})}{E\!-\!E_0\!-\!\frac{{q'}^2}{2m}}
 {\rm d}{\bf q}\,{\rm d}{\bf q'},
 \label{me3bf}
\end{equation}
where $t_d$ is the isospin of the bag (we remind that $S_d=J_i,\,
s_d+t_d=1$). For brevity, we omitted here the labels of the basis functions. After partial-wave 
decomposition  (cf. 
eq.(\ref{ovfull}))
\begin{equation}
\frac{\chi^{J_iL_i}({\bf
q})}{E-E_o-q^2/2m}=\sum_{\lambda}\psi^{J_iL_i}_{\lambda}(q)
Y_{\lambda}(\hat{\bf q}),
 \label{pwdec}
\end{equation}
 integral (\ref{me3bf}) reduces to a sum of integrals of the form
\begin{equation}
  \langle V_{\rm Coul}^p \rangle^{\lambda} =
  \int \psi_{\lambda}(q)V_{cp}^{\lambda}(q,q')\psi'_{\lambda}(q')
  \,q^2{\rm d}q\,{q'}^2{\rm d}q',
 \label{mevplam}
\end{equation}
where
 \begin{equation} 
 \fl V_{cp}^{\lambda}(q,q') = \frac{2}{\pi}\int j_{\lambda}(qr)\,
 \frac{1}{r} \, j_{\lambda}(q'r) \,r^2{\rm d}r 
 = \frac{2}{\pi}\frac{1}{2qq'}Q_{\lambda}(z);\qquad z=\frac{q^2+{q'}^2}{2qq'}.
 \label{vplam}
\end{equation}
Here, $Q_{\lambda}(z)$ is the Legendre function of the second kind. Now, we will replace the
Coulomb  potential $1/r$ between the point-like charges in eqs.(\ref{mevplam}-\ref{vplam})
with the  corresponding potential between  the ``smeared'' charges:
\begin{equation}
 \fl  \langle V_{\rm Coul}(\alpha)\rangle^{\lambda} =
  \int \psi_{\lambda}(q)\left [
  \frac{2}{\pi}\int j_{\lambda}(qr)
 \frac{{\rm erf}(r\sqrt{\alpha})}{r} j_{\lambda}(q'r) \,r^2{\rm d}r \right ]
  \psi'_{\lambda}(q')  \,q^2{\rm d}q\,{q'}^2{\rm d}q'.
 \label{mesmear}
\end{equation}

It is necessary to comment the calculations of such integrals. In the momentum representation, the 
integrals include the Coulomb singularity. Thus, they must be carefully integrated numerically. In 
practice, it is much more convenient to treat them in the coordinate space especially in the case
of  the ``smeared'' Coulomb interaction. However, the presence of the propagators
$(E-E_0-q^2/2m)^{-1}$ in  our case requires the use of the momentum representation from the
beginning. Hence, we calculated  the above Coulomb integrals as follows. Taking into account that the overlap
functions $\chi(\bf q)$ in the  Gaussian basis reduce to a sum of Gaussians, we approximated the
above propagators by a sum of few Gaussians. Then, $\psi_{\lambda}(q)$ entering into
eq.(\ref{pwdec}) takes  the form
\begin{equation}
   \psi_{\lambda}(q)=\sum_{i}\sum_{l=\lambda}^{\lambda +n} C^i_{\lambda l}
   \,q^l\,{\rm e}^{-\beta^i_lq^2},
 \label{psigaus}
\end{equation}
where the additional degrees of $q$ arise due to the use of an antisymmetrized basis. Now, the 
integral (\ref{mesmear})
reduces to a sum of terms involving only one-dimensional integrals:
\begin{equation}
 \fl  I_{\mu}(\beta,\alpha)=\int \frac{{\rm erf}(r\sqrt{\alpha})}{r}\,r^{2+2\mu}
   \,{\rm e}^{-\beta r^2} {\rm d}r
 = \frac{1}{2}\frac{1}{\alpha^{\mu +1}}\sum_{k=1}^{\mu}
   \frac{k!}{(\frac{\beta}{\alpha})^{k+1}}\,
   \frac{(2\mu -2k+1)!!}{2^{\mu -k}(\frac{\beta}{\alpha}+1)^{\mu-k+1/2}},
 \label{Imu}
\end{equation}
where $C^k_{\mu}$ are the binomial coefficients.
Thus, this technique reduces the whole calculation of the three-body Coulomb interaction matrix to 
completely analytical formulas, which considerably simplify the variational calculation.

\subsection {Rms matter and charge radii}

In the $3N$ channel, the rms radii of the proton ($r_p$) and neutron ($r_n$) distributions are
defined  by the standard way:
\[r_p^2 = \frac{3}{N_p}\langle \Psi_{3N}|\frac{1+\tau_3^{(1)}}{2}
 \,r_1^2 |\Psi_{3N}\rangle,\]
 \begin{equation}
 r_n^2 = \frac{3}{N_n}\langle \Psi_{3N}|\frac{1-\tau_3^{(1)}}{2}
\,r_1^2 |\Psi_{3N}\rangle,
 \label{rpn}
\end{equation}
where $r_1=(2/3)\rho_1$ is the distance between particle (1) and the system  
center of mass,   
 $N_p=3/2+T_z$ and $N_n=3/2-T_z$ are the numbers of protons and neutrons, 
 respectively. $N_p$ is equal to the total charge of the system $Z$.
Then, the  rms matter radius is equal to
\begin{equation}
 r_m^2 = \frac{N_pr_p^2 + N_nr_n^2}{3}=\langle\Psi_{3N}|r_1^2 |\Psi_{3N}\rangle
 \label{rmatt}
\end{equation}
The rms charge radius in the $3N$ sector is also defined conventionally:
\begin{equation}
 \langle r_{ch}^2\rangle_{3N} = r_p^2+R_p^2-\frac{N_n}{N_p}R_n^2,
 \label{rch3N}
\end{equation}
where $R^2_p=0.7569$~fm$^2$ and $R_n^2=-0.1161$~fm$^2$ are the squared charge radii of the proton 
and neutron, respectively. 

Further, we define the rms charge radius in $6qN$ channels as
\begin{eqnarray}
  \nonumber \langle r_{ch}^2\rangle_{6qN} = \frac{1}{ZP^{(1)}_{6qN}}
 \left\langle\Psi_{6qN}^{(1)}\right |\frac{1+\tau^{(1)}_3}{2}(r_1^2 + R_p^2)
 + \frac{1-\tau^{(1)}_3}{2}R_n^2\\
  +(1+\hat{t}^d_3)r_d^2+ 
 \sum_{t,t_z}\Gamma_{t,t_z}R_d^2(t,t_z)\left |\Psi_{6qN}^{(1)}\right\rangle,
 \label{rch6qN}
\end{eqnarray}
where
\[ {\bf r}_1=\frac{m_d}{m_d+m_N}{\b \rho_1} \equiv \alpha{\bf \rho_1} 
\mbox{ and }{\bf r}_d=-\frac{m_N}{m_d+m_N}{\b \rho}_1 = -(1-\alpha){\b \rho}_1\] 
are the coordinates of the third nucleon (with number 1) and the bag 
in the c.m.s., $\hat{t}^d_3$ is the operator of
the third component of  the bag isospin, $\Gamma_{t,t_z}$ is the projector into 
isospin state of the bag with definite values of its isospin $t$ and
$z$-projection $t_z$. The $R_d(t,t_z)$ is the value of charge radius of the
bag in the specific isospin state. These values, in general, are different, but 
their difference should be rather small and,thus, can be safely ignored in subsequent
calculations keeping in mind the relatively low probability of the bag-component.Thus
we take value $R_d=0.6$~fm for the mean charge  radius of the bag in all 
isospin states, except the state with $t_z=-1$ corresponding to ($nn$)-bag. 
For latter (uncharged) state we put $R^2_d(1,-1)=0$.

The total rms charge radius of the $3N$ system with allowance for both 
the three-nucleon and dibaryon-nucleon components thus takes 
the form
\begin{equation}
  \langle r_{ch}^2\rangle=(1-P_{6qN}) \langle r_{ch}^2\rangle_{3N} +
  P_{6qN} \langle r_{ch}^2\rangle_{6qN}.
   \label{rchtot}
 \end{equation}

Similarly to eq.~(\ref{rpn}), one can define the rms radius of the proton (neutron) distribution 
 in the $6qN$ channel as 
\begin{equation}
  \langle r_{\{{p\atop n}\}}^2\rangle_{6qN}=  \frac{\langle \Psi_{6qN}^{(1)}|
  \frac{1}{2}(1\pm\tau_3) r_1^2 |\Psi_{6qN}^{(1)}\rangle }
 {\langle \Psi_{6qN}^{(1)}|
  \frac{1}{2}(1\pm\tau_3)  |\Psi_{6qN}^{(1)}\rangle },
   \label{rp6qN}
 \end{equation}
where the denominator determines  the average number of protons $N^{6qN}_p$ 
(or neutrons $N^{6qN}_n$) in the $6qN$ channel: 
\begin{equation}
 N^{6qN}_{\{{p\atop n}\}} = \frac{\langle \Psi_{6qN}^{(1)}|
  \frac{1}{2}(1\pm\tau_3)  |\Psi_{6qN}^{(1)}\rangle}{P^{(1)}_{6qN}} .
   \label{np6qN}
 \end{equation}
 One can note that, if to neglect the difference between isosinglet and
 isotriplet wavefunctions in the $6qN$ channel, then these numbers are equal 
 $N_p=1/3$ and 2/3  ($N_n=2/3$ and 1/3) for $^3$H and $^3$He respectively. 

The total rms radii of the nucleon distributions including both the 
$3N$ and three $6qN$ components are
\begin{equation}
  \langle r_{\{{p\atop n}\}}^2\rangle=
  (1-P_{6qN}) r_{\{{p\atop n}\}}^2\ +
  P_{6qN} \langle r_{\{{p\atop n}\}}^2\rangle_{6qN}.
   \label{rntot}
 \end{equation}

\subsection{Role of the $pn$ mass difference}

To accomplish calculation for the accurate Coulomb displacement energy $\Delta E_C=E_B(^3{\rm H})-
E_B(^3{\rm He})$, one should take into consideration some tiny effects associated with the mass 
difference between the proton and neutron. It is well known~\cite{Gloeckle} that the above mass 
difference makes rather small contribution to the difference between $^3$He and $^3$H binding 
energies. Therefore, it is usually taken into account in the perturbation approach. However, since the 
average kinetic energy in our case is twice the energy in conventional force models, this correction 
is expected to be also much larger in our case. Hence, we present here the explicit exact evaluation for 
such a correction term without usage of the perturbation theory.

It should be added that in our model involving various $6qN$ components the similar effect 
originated from the mass difference of dibaryons in different charge states with $Z=0,1$ and 2 
should also be taken into 
account. This mass difference is equal about $\Delta M_d^C \sim 3$~MeV.~\footnote{%
The mass difference between baryons with different $ST$ values is already included in our force 
model.} The latter effect seems to yield a negligible correction to the $\Delta E_C$ value, because 
the total probability of all $6qN$ components does not exceed 10 -- 11~\%, while the nucleon mass 
difference is half the $\Delta M_d^C$ value, i.e. ca. 1.5~MeV, at the probability of the $3N$ channel 
ca. 90\%.

In the conventional isospin formalism, one can consider that the $^3$H and $^3$He nuclei consist of 
the equal-mass nucleons: 
\[{m}=\frac{m_p+m_n}{2},\]
so that $m_p= m+\Delta m/2, \; m_n=m-\Delta m/2$, where $\Delta m= m_p-m_n $. The simplest way to 
include the correction due to 
the mass difference $\Delta m$ is to assume that all particles in $^3$H have the average mass 
\[\bar{m}_H =\frac{2m_n + m_p}{3}=m-\frac{1}{6}\Delta m,\]
while they have the different average mass 
\[\bar{m}_{He} =\frac{2m_p + m_n}{3}=m+\frac{1}{6}\Delta m \] 
in $^3$He. In spite of smallness of parameter $\Delta m/m$, the perturbation theory in this parameter
does not work. So we used the average mass 
$\bar{m}_H$ in calculation of $^3$H and $\bar{m}_{He}$ in calculation of $^3$He.
The results of these corrections are given in fifth row of Table~2.

\section{Results of calculations}

Here, we present the results of the $3N$ bound-state calculations based on two variants of the DBM.
\begin{itemize}
\item [(i)]
In first version developed in~\cite{IJMP}, the dressed-bag propagator includes three loops, two of 
them are of the type shown in Fig.2 of ref.~\cite{IJMP}, in which each loop was found with the $^3P_0$ model. 
The third loop consists of two such vertices and a convolution of the $\sigma$-meson and $6q$-bag 
propagators (see the Fig.~2 in ref~\cite{IJMP}).
\item[(ii)] In the second version, we replaced two above  loops with the effective 
Gaussian form factor $B(k)$, which describes the direct $NN \to 6q+\sigma$ transition, i.e., 
the direct transition from the $NN$
channel to the dressed-dibaryon channel.
\end{itemize}

Both versions have been fitted to the $NN$ phase shifts in low partial waves
up to an energy of 1~GeV  with almost the same quality. Therefore, they can be
considered on equal footing. However, version  (ii) has one important
advantage. Here, the energy dependence arising from the convolution of the
two  propagators involved into the loop, i.e., the propagators of the
$\sigma$-meson and bare dibaryon,  describes (with no further correction) just
the energy dependence of the effective strength of the  $NN$ potential
$\lambda^{(2)} (E)$, which is thereby {\em taken directly from the above loop
integral}. In contrast, in the first version of the model, two additional
$qq\pi\pi\sigma$ loops give a rather  singular three-dimensional integral
for $\lambda^{(1)}(E)$, where the energy dependence at higher  energies should
be corrected by a linear term. 
The main difference between the results for both versions is that the energy
dependence of  $\lambda(E)$ for the second version is much weaker than that
for the first variant. In addition, this  energy dependence leads both to the
decrease in the contribution of the $6qN$ component to all $3N$  observables
and thus to the increase in the contribution of the two-body force as compared
to the three-body force. 

Table~1 presents the calculation results for the two
above versions for the following characteristics: -- the weights of the $6qN$
channels and $D$ wave in the total $3N$ function, as well as the weight of the
mixed-symmetry $S'$ component (only for the $3N$ channel); -- the averaged individual
contributions from the kinetic energy $T$, two-body interactions $V^{(2N)}$ 
plus the kinetic energy $T$ and three-body force ($V^{(3N)}$) due to one-sigma
and two-sigma exchanges  to the total Hamiltonian expectation.

\begin{table}[h]
\caption{Results of the $3N$ calculations with two- and three-body forces for two variants of 
the dressed-bag model.}
\begin{tabular}{|*{8}{c|}}  \hline
  & $E$& $P_D$ \% & $P_{S'}$ \%& $P_{6qN}$ \%
  & \multicolumn{3}{|c|}{Individual contributions to $H$, MeV} \\
  \cline{6-8}
        & MeV&&&                      &$T$& $T+V^{(2N)}$&$V^{(3N)}$ \\
 \hline
 \multicolumn{8}{|c|}{ $^3$H}\\ \hline
 DBM(I) $g=9.577^{*)}$  & -8.482 & 6.87 & 0.67 & 10.99 & 112.8 & -1.33 & -7.15  \\
 \hline
  DBM(II) $g=8.673^{*)}$& -8.481 & 7.08 & 0.68 & 7.39 & 112.4 & -3.79 & -4.69   \\
 \hline
 AV18+UIX$^{1)}$ & -8.48& 9.3 & 1.05 &  -   & 51.4 & -7.27 & -1.19   \\
 \hline \hline
\multicolumn{8}{|c|}{ $^3$He}\\ \hline
 DBM(I) & -7.772  & 6.85 & 0.74 & 10.80 & 110.2 & -0.90 & -6.88   \\
  \hline
 DBM(II) & -7.789  & 7.06 & 0.75 & 7.26  & 109.9 & -3.28 & -4.51   \\
 \hline
 AV18+UIX$^{1)}$ & -7.76 & 9.25 & 1.24 & -     & 50.6 & -6.54 & -1.17   \\
 \hline \hline
\end{tabular}\\[0.5 mm]

{\small \noindent $^{*)}$ These values of $\sigma NN$ coupling constant in 
$^3$H calculations have been chosen to reproduce the exact binding energy 
of $^3$H nucleus. The calculations for $^3$He have been
carried out without any free parameters.\\
\noindent $^{1)}$ The values are taken from~\cite{Pieper2001}.
} 
\end{table}

To compare with the respective results for the conventional $NN$ potential
models, Table~1  also  presents the results of recent calculations with the
Argonne potential AV18 and Urbanna-Illinois  three-body force
UIX~\cite{Pieper}. The Coulomb displacement energies $\Delta E_C$, together
with the individual contributions to the  $\Delta E_C$-value, are presented in
Table~2. The rms radii of the charge and proton distributions in $^3$H and
$^3$He found in the
impulse approximation, as well as  the respective experimental values  and
results obtained for AV18 + UIX $NN$ forces, are presented in Table~3. To
demonstrate the separate  contributions of the three-nucleon and
dibaryon-nucleon channels to these observables, we also present  the values
calculated separately with only nucleonic and $6qN$ parts of the total
wavefunction.

\begin{table}
\caption{Contribution of various terms (in keV) of the
interaction to the $^3$H - $^3$He mass difference.}
\begin{tabular}{|l|r|r|r|}  \hline
Contribution              & DBM(I)& DBM(II) & AV18+UIX \\ \hline 
point Coulomb 3N only    & 598    & 630 & 677 \\ \hline 
point Coulomb 3N+6qN     & 840    & 782 & -   \\ \hline 
smeared Coulomb 3N  only & 547    & 579 & 648 \\ \hline 
smeared Coulomb 3N+6qN   & 710    & 692 & -   \\ \hline 
$np$ mass difference     & 46     & 45  & 14  \\ \hline 
nuclear CSB$^{1)}$       & 0      & 0   &+65  \\ \hline 
magnetic moments  \&     & 17     & 17  & 17   
\\ spin-orbit$^{2)}$ &&& \\ 
\hline Total             & 773    & 754 & 754 \\ \hline

\end{tabular}\\[0.5 mm]

{\small \noindent $^{1)}$ See Table~4.

\noindent $^{2)}$ Here we use the value for this correction from 
ref.~\cite{Gloeckle} 
}
\end{table}

\begin{table}
\caption { Rms proton, $r_p$, and charge, $r_{\rm ch}$, radii (in fm) in DBM
approach}

\begin{tabular}{|*{6}{c|}}  \hline
 \multicolumn{2}{|c|}{model} &\multicolumn{2}{|c|}{$^3$H}&
 \multicolumn{2}{|c|}{$^3$He} \\ \cline{3-6}
 \multicolumn{2}{|c|}{ }& $r_p$ & $r_{\rm ch}$  & $r_p$ & $r_{\rm ch}$  \\ \hline
        & 3N  & 1.625  & 1.779  & 1.805 & 1.989    \\ \cline{2-6}
DBM(I)  & 6qN & 1.608  & 1.188  & 1.854 & 1.412    \\ \cline{2-6}
      & total & 1.625  & 1.724  & 1.807 & 1.935    \\ \hline
        & 3N  & 1.613  & 1.769  & 1.795 & 1.980    \\ \cline{2-6}
DBM(II)  & 6qN & 1.573  & 1.171  & 1.829 & 1.396   \\ \cline{2-6}
      & total & 1.613  & 1.732  & 1.796 & 1.944    \\ \hline
 \multicolumn{2}{|c|}{AV18 + UIX} 
              & 1.59$^{*)}$   &        &  1.76$^{*)}$ &          \\ \hline
 \multicolumn{2}{|c|}{Experiment}   
              & 1.60$^{*)}$   & 1.755  &  1.77$^{*)}$ & 1.95     \\ \hline

\end{tabular}\\[0.5 mm]

{\small 
\noindent $^{*)}$ These values are taken from~\cite{Pieper2001}. The ``experimental'' 
values of point proton radii $r_p$ have been obtain there from charge radii by removing the proton
and neutron charge radii  0.743~fm$^2$ and -0.116~fm$^2$ respectively.
} 

\end{table}

We present here a few comments concerning the results in Tables~1-3. 

{\em Comments to Table~1.}
\begin{itemize} 
\parindent=2em 
\item[(i)] It is seen that an admixture of the mixed-symmetry
$S'$ component in our $3N$ wavefunction is almost half that for the
conventional force model (e.g., AV18+UIX). This difference can be attributed to
the fact that the relative contribution of two-body interactions  to the total
$3N$ binding energy in our approach is much less than that in the conventional
force  models 
(this follows from the results presented in the seventh and eighth columns
of Table~1).  While it is well known that the weight of the $S'$ component is
proportional to the difference between two-body spin-singlet and spin-triplet
$NN$ interactions, the leading contribution of three-body forces in our
approach comes from the scalar-isoscalar 3BF that is completely insensitive to
the above difference. Therefore, as a result of this redistribution of various
force components, the weight of the $S'$ component decreases by almost half. A
similar but weaker diminution with respect to conventional force models is also
seen in the weight of the $D$-wave component $P_D$. This decrease has the same
origin, viz. the suppression of the two-body force contribution and the large
increase in the scalar three-body force contribution.

\item[(ii)]  Another remarkable distinction from the conventional force models is the large
increase in the average kinetic (and potential) energy, viz. 112~keV vs.
 50~keV in conventional force models (see the sixth column in Table~1). This
increase is caused by the appearance of the short-range radial nodes and
respective loops in the radial $3N$ wavefunctions (see Fig.~5 in the preceding
paper). This large enhancement in nucleon velocities will strengthen all the effects
associated with the nucleon currents, relativistic effects, meson-exchange
contributions to electromagnetic observables, etc.
\end{itemize}
 
{\em Comments to Table~2.}  Here, we emphasize three important points.
\begin{itemize} 
\item[(iii)] First, it is seen quite a large contribution from the Coulomb three-body force
(cf. the differences between the entries ``Coulomb $3N$ only'' and `` Coulomb
$3N+6qN$'' in this table). The second and third rows correspond to the Coulomb
interaction between point-like charges, while the fourth and fifth rows include
results for the Coulomb interaction between properly smeared charge
distributions. In both cases, the contribution of the three-body Coulomb
interaction (which has been completely overlooked in previous works) is as
large as ca. 110 -- 240~keV and, along with other minor effects, can
quantitatively explain the Coulomb displacement energy of $^3$He.

\item[(iv)] The second point, which is closely interrelated to the first one,  is rather
high sensitivity of all above Coulomb contributions to the smearing of the
charges (both for the proton and $6q$ dibaryon) with the Gaussian distribution.
Table~2 shows that the inclusion of the smeared charge distribution reduces 
the Coulomb two-nucleon force contribution by 51~keV for both versions of the
model. It should be compared to a difference of 29~keV in the Coulomb
interactions between point-like and smeared nucleons  for the AV18 + UIX
force model. Smearing the charges leads also to significant reducing of the
three-body Coulomb contribution: from 242 to 163~keV for version~I and  from
152 to 113~keV for version~II.  However, even in the  minimum scenario, we
obtain an additional contribution of 113~keV from the three-body (smeared)
Coulomb force.

\item[(v)] The third interesting feature, which is distinguished from the conventional
model result, is a quite large effect of the (small) $np$ mass difference on
the $3N$ Coulomb displacement energy. This effect is about twice the respective
contribution for the AV18 + UIX force model. This enhancement is attributed to
the much increased average kinetic energy in our approach. Thus, the variation
of this energy due to the $np$ mass difference should be also much larger.

We also add a small correction due to electromagnetic interactions and
spin-orbit electromagnetic interaction (as in conventional models, we take a
value of 17~keV for this correction). Including all these corrections, we
obtain the total value $\Delta E_C^I=773$~keV for version~I  and $\Delta
E_C^{II}=754$~keV for version~II. Thus, we found a quite small space for
nuclear CSB effects: -9~keV for DMB(I) and +10~keV
for DBM(II). These values should be compared to a significant value of 65~keV
for the AV18 + UIX force model. The more detailed discussion of CSB effects in
our approach in the next section further corroborates this important
conclusion: the admissible value of CSB effects in the DBM is noticeably
smaller than that in the conventional force models.
\end{itemize}
 
{\em Comments to Table~3.}  
\begin{itemize} 
\item[(vi)]  
The rms radii of the proton and charge distributions are
presented in Table~3 for two versions of our model in comparison with the
results for the A18 + UIX force model. As is seen in Table~3, the  rms charge
radii for the $6qN$ component in both $^3$H and $^3$He are much smaller than
those for the $3N$ component (as could be expected in advance). On the other
hand, the rms charge radii for the $3N$ component turn out to be larger than
the respective experimental values in both $^3$H and $^3$He. Thus, it is the
contribution of the $6qN$ component to the total wavefunction of the $3N$
system that provides quite good agreement of the rms charge radii with the
respective experimental values. 

 A small underestimation of charge radii (especially for $^3$H) can be due to too 
small value for charge radius 
of the bag (0.6~fm) accepted in our calculations. This value is, in fact, the 
quark-core radius of the bag, but our estimate shows that the pion cloud will 
increase its charge radius up to $0.65\div 0.68$~fm. This will lead to increase of 
charge radius of $6qN$ component and, therefore, to some increase of the total charge radius.
Besides that,there is a contribution to charge radius from the model-dependent 
two-nucleon current operator. As it shown in ref.~\cite{Marc}, this contribution for 
AV18+UIX force model is about 0.014~fm for $^3$H and 0.009~fm 
for $^3$He. 
  
\end{itemize}

\section{Discussion}
Here, we will discuss the main results found in the work in the general context of few-body physics 
and compare them with the respective results based on conventional force models. Particular attention 
will be paid to some general conclusions that can be derived from the results presented here.
Let us begin with the results for the Coulomb displacement energy $\Delta E_C=E_B({}^3{\rm H}) -
E_B({}^3{\rm He})$. We emphasize three important points, where our results differ from those for 
conventional models.
\begin{itemize}

\item[(i)] First, we found a serious difference between conventional and our
approaches in the short-range  behaviour of wavefunctions even for the nucleon
channel. Conventional $3N$ wavefunctions are strongly  suppressed along all
three interparticle coordinates $r_{ij}$ due to the short-range local
repulsive  core, while our wavefunctions (in the $3N$ channel) have stationary
nodes and short-range loops along  both all $r_{ij}$ and the third Jacobi
coordinates $\rho_k$. Such a node along the $\rho$ coordinate presents also
in  the $6qN$ relative-motion wavefunction (see Fig.5 in ref.~\cite{Ku03}). This very
peculiar short-range  behaviour of our wavefunctions leads to a strong
enhancement of the high-momentum components of  nuclear wavefunctions, which
is indicated by various modern experiments, e.g., $^3{\rm 
He}(e,e'pp)$~\cite{Hepp} or $pp \to pp\gamma$ etc. where high momentum
transfers appear. On the other hand, these short-range radial loops lead to
significant errors when using the Coulomb  interaction between point-like
particles within our approach. Hence, we must take into account {\em the
finite radii of charge distributions in the proton and $6q$  bag}. Otherwise,
all Coulomb energies are overestimated.

\item[(ii)] Another important effect following from our calculations is a
quite significant contribution of the  $6qN$ component to $\Delta E_C$. In
fact, just this interaction, which is completely missing in  conventional
nuclear force models, makes the main contribution (ca. 100 keV) to filling the
gap in  $\Delta E_C$ between conventional $3N$ calculations and experiment.

\par The large magnitude of this $3N$ Coulomb force contribution is explained by
two factors: first, a rather short average distance $\langle\rho^2\rangle$
between the $6q$ bag and third nucleon (which  enhances the Coulomb
interaction in the $6qN$ phase) and, second, a significant weight of the 
$6qN$ components where the bag has the charge  +1 (i.e., it is constructed
from an $np$ pair). This  specific Coulomb repulsion in the $6qN$ channel
should appear also in all other nuclei where the total  weight of such
components is about 10\% and higher. Therefore, it should strongly contribute
to the  Coulomb displacement energies over the entire periodic table and could
somehow explain the long-term  Nollen-Schiffer paradox~\cite{Nollen} in this
way.

\item[(iii)]  The third specific effect that has been found in this study and
contributes to the  quantitative explanation of $\Delta E_C$ is a strong
increase in the average kinetic energy $\langle  T\rangle$ of the system. This
increase in $\langle T\rangle$ has been already discovered in the first  early
$3N$ calculations with the Moscow $NN$ potential model~\cite{Mosc} and results
in a similar  nodal wavefunction behaviour along all interparticle coordinates
but without any non-nucleonic  component.

\par The increase in $\langle T\rangle$ leads to the proportional increase in
the $np$ mass difference  correction to $\Delta E_C$. As is seen in Table~2,
this correction in our case is not very small and  contributes significantly
to $\Delta E_C$. Many other effects attributed to increasing the average 
kinetic energy of the system will arise in our approach, e.g., numerous
effects associated with the  Fermi motion of nucleons in nuclei.

 It is worth also to estimate here the possible contribution of $9q$-bag
component in $^3$He to the $\Delta E_C$, which has been omitted in our present 
calculations. If one assumes that the radius of the $9q$-bag is near to that 
for $6q$-system, i.e. $r_{9q} \sim 0.6$~fm, the Coulomb energy of this bag 
(with the charge $Z=2$) $Ze^2/r_{9q}$ is about 3~MeV. The probability of the 
$9q$-bag is 
expected to be around 0.1 -- 0.2~\%. So that the additional Coulomb 
energy contribution to $\Delta E_C$ should be about $3\cdot (0.1\div 0.2)\cdot
10^{-3}\sim 
3\div 6$~keV, i.e. a quite small value as compared to the quantity 120~keV 
coming from $6qN$ component. It can also be expected that the corrections to 
kinetic energy difference due to contribution of $9q$ component would be very 
small. 

\end{itemize}

The best explanation for the $\Delta E_C$ value in the framework of
conventional force models  published up to date~\cite{Gloeckle} is based on
the introduction of some CSB effect, i.e., the  difference between $nn$ and
$pp$ strong interactions. At present, two alternative values of the $nn$ 
scattering length are assumed:
\begin{equation}
a^{(1)}_{nn}= -18.7 \mbox{ fm, and } a^{(2)}_{nn}=-16.3 \mbox{ fm}.
\label{ann}
\end{equation}
The first value has been extracted from the previous analysis of experiments $d(\pi^-
,\gamma)nn$~\cite{dpigam} (see also ref.~\cite{dpig1} and refs. therein) and is used in all current 
$NN$ potential models, while the second value in (\ref{ann}) has been derived from numerous
three-body  breakup experiments $n+d \to nnp$ done for the last three decades. In recent years, such
breakup  experiments are usually treated in the complete Faddeev formalism, which includes most
accurately both  two-body and three-body forces~\cite{a16}. Thus, this $a_{nn}$ value is considered
as quite reliable.  However, the quantitative explanation for the $\Delta E_C$ value in conventional
force models uses  just the first value of $a_{nn}$ as an essential point of all the construction.
At the same time, the  use of the second value $a_{nn}(=-16.3\mbox{ fm})$ (which is not less
reliable than the first one)  fails completely the above explanation!

Therefore, in order to understand the situation more deeply and to determine
the degree of sensitivity  of our prediction for $\Delta E_C$ to variation in
$a_{nn}$,  we made also our $3N$ calculations with  two possible values of
$a_{nn}$ from eq.(\ref{ann}).  These exact calculations have been carried out with
the effective values of the  singlet-channel coupling constant corresponding
to the $V_{NqN}$ part of the  $NN$ force:

\begin{equation}
 \lambda^{\rm eff}_{^3He}(^1S_0)=\frac{1}{3}\lambda_{pp}+
 \frac{2}{3}\lambda_{np};
 \label{effconstHe}
 \end{equation}
\begin{equation}
 \lambda^{\rm eff}_{^3H}(^1S_0)=\frac{1}{3}\lambda_{nn}+
 \frac{2}{3}\lambda_{np}.
 \label{effconstH}
 \end{equation}
  In the above calculations, we use the value $\lambda_{np}=328.9$~MeV that 
provides the accurate description of the $^1S_0 $ $np$ phase shifts and the
experimental  value of the $np$ scattering length
$a_{np}=-23.74$~fm~\cite{IJMP}. Here, we employ the value
$\lambda_{pp}=325.523$~MeV fitted to the well-known experimental magnitude
$a_{pp}=-8.72$~fm and two $\lambda_{nn}$ values corresponding to two available
alternative values of the $nn$ scattering length: $a^{(1)}_{nn}=-16.3$~fm and
$a^{(2)}_{nn}=-18.9$~fm. The calculation results are presented in Table~4.

\begin{table}
 \caption{Contribution of charge symmetry breaking effects to the $^3$H -
$^3$He mass
 difference.}
 
 \begin{tabular}{|c|c|c|}\hline
 & \multicolumn{2}{|c|}{$\Delta E_C$, keV }\\ \cline{2-3}
 $a_{nn}$, fm & DBM(I) & DBM(II)\\ \hline
 -16.3 & -18   & -39 \\ \hline
 -18.9 & +45   & +26 \\ \hline
 \end{tabular}

\end{table}

As is seen in Tables 2 and 4, within the DBM (version I) one has $\Delta E_C$=
773-18=754 keV, so that the version I of DBM can reasonably reproduce the
the experimental Coulomb displacement energy $\Delta E_C$ with the lower 
(in modulus) value
$a_{nn}=-16.3$~fm, while this model overestimates  $\Delta E_C$ by 54~keV with
the  larger (in modulus) value $a_{nn}=-18.9$~fm. Thus, the DBM approach, in
contrast  to the conventional force models, prefers the lower (in modulus)
possible value  -16.3~fm of the $nn$ scattering length. 

Now, let us discuss shortly the magnitude of CSB effects in our model.   The
difference between $a_{nn}$ and so-called ``pure nuclear'' $pp$  scattering
length $a_{pp}^N$ is usually considered as the measure of CSB effects at low
energies. The value $a_{pp}^N$ is extracted  from $pp$ scattering data when the
Coulomb potential is disregarded. The model dependence of the latter quantity
was actively  discussed in the 1980s~\cite{SauWall,Rahman,Albev}. However, the
majority of modern  $NN$ potentials fitted to the experimental value
$a_{pp}=-8.72$~fm give the  value $a^N_{pp}=-17.3$~fm when the Coulomb
interaction is discarded. It is the  value that is adopted now as an
``empirical'' value of the $pp$ scattering  length~\cite{Machl}. Thus, the
difference between this value and $a_{nn}$ is  usually considered as the
measure of CSB effects. However, our model (also  fitted to the experimental
value $a_{pp}=-8.72$~fm) gives a quite surprising  result for pp-scattering
length when the Coulomb effects are removed:

\begin{equation}
 a^N_{pp}(DBM) = -16.57 \mbox{ fm},
 \label{anpp}
 \end{equation}
 which differs significantly from the above conventional value (by 0.8~fm) due
 to the explicit energy dependence of the $NN$ force in our approach.
 
 Thus, if the difference $a^N_{pp}-a_{nn}$ is still taken as the measure of 
 CSB effects, the smallness of this difference obtained in our model testifies 
to a small magnitude of the CSB effects, which is remarkably smaller than the 
values derived from conventional OBE models for the $NN$ force.
 
Now, let us pass to the data from Table~3 for the radii of the charge and
proton distribution in $^3$H  and $^3$He. It is seen that both our versions
(DBM(I) and DBM(II)) give quite similar values for all  radii. The most
interesting point here is the importance of $6qN$ component contributions. In
fact,  the contribution of the $6qN$ channel shifts all radii, i.e., $r_{ch}$
and $r_p$ in $^3$H and $^3$He,  predicted with pure nucleonic components, much
closer to the respective experimental values. 

Thus, the dibaryon-nucleon component also works in a right way in this aspect.
It is interesting to  note that, in general, the predictions of our {\em
two-phase model} are quite close to those of the  conventional single-phase
AV18 + UIX model. This means that (at least for many static characteristics) 
our multi-channel model is effectively similar to a conventional purely
nucleonic model. However, this  similarity will surely hold only for the
characteristics that are sensitive mainly to low momentum  transfers, while the
properties and processes involving high momentum transfers will be treated in
two  alternative approaches in completely different ways.
 
  It is worth here to add a few remarks about role of the 
Coulomb effects in $3N$- and $4N$-continuum and its interpretation on the basis 
of the present model. There are a few long-standing puzzles in the field 
which still cannot be resolved within conventional models of $2N$ and $3N$ 
forces. Here are:
\begin {itemize}
 \item[(i)] The values of $A_y$ and $T_{11}$ in $nd$ and $pd$ radiation 
    capture at low energies are strongly underpredicted by 
    conventional theoretical approach~\cite{Wulf}.
 \item[(ii)] The pronounced discrepancy for the fore-aft asymmetries in mirror 
reactions $^3{\rm H}(\gamma,n){}^2{\rm H}$ and 
$^3{\rm He}(\gamma,p){}^2{\rm H}$ at energies a few MeV above the 
thresholds and also for inverse capture processes~\cite{Mitev,Skopic}.
 \item[(iii)] The ratio of cross sections $(\gamma,p)$-to-$(\gamma,n)$ for 
 $^4$He at 
energies a few MeV above the thresholds is as large as 1.7 while the 
conventional theory predicts only the values ca. 1.3 - 1.4~\cite{Ward}.
\end{itemize}

 The discrepancy in (i) can be reduced somehow by inclusion of the strong 
two-body meson-exchange currents incorporating $\Delta$-current with 
(unnaturally) high cut-off parameter $\Lambda_{\pi N\Delta}$ and 
$\Lambda_{\pi N\Delta}$\cite{Carlson,Yaf}.
 The puzzle in (iii) could be explained or reduced by assuming a strong 
charge-symmetry breaking force component which is in an evident 
contradiction with results of other experiments and also with conclusions 
of the present force model.
 
 It should be stressed that all three above-mentioned puzzles are  interrelated
to contribution of $P$-states in $nd$ and especially  $nT$ systems. For
example, as has been found in recent $4N$ calculations~\cite{Cies}   the
$P$-wave peak in $nT$ elastic scattering at energies ca. 3 - 4 MeV cannot  be
explained by the fully realistic 4N-calculations within the conventional  force
model. In addition, the issue (iii) can be explained by an enhancement  of the
Coulomb effects in $p\!-\!{}^3$H exit channel.

 As follows from the results of the present and preceding works, our force 
model predicts inevitably an enhancement of the $P$-wave contributions in  $Nd$
and especially $NT$ near continuum (due to additional strong scalar 
$3N$-force) and also an enhancement of the Coulomb effects in $pd$ and $p\,^3$H
near continuum. So the present approach could remove or reduce noticeably the
above discrepancies in $3N$ and $4N$ low-energy continuum.

\section{Conclusion}
Here, we will summarize the main results of this work. In the previous work, we
fixed the only  coupling constant, $g_{\sigma NN}$, to obtain the experimental
value of the triton binding energy.  Then, all other calculations in both
previous and this works did not include any fitting parameter.  Thus, their
results can be considered as a stringent test for the proposed new model for
$2N$ and $3N$  forces.

First, we point to the precise value obtained for the Coulomb displacement
energy $\Delta E_C$ of the  $A=3$ system in the developed model. It should be
emphasized that, contrary to other studies based on  conventional force models
(using the $2N$ and $3N$ forces generated via the meson-exchange mechanism), 
this explanation {\em does not require any noticeable CSB effect}, although our
model is still  compatible with such effects. However, these CSB effects do not
contribute remarkably to $\Delta E_C$  in our approach. Two basic sources of
this contribution, which differ from conventional force models,  should be
indicated here:
\begin{itemize}
\item[--] the three-body Coulomb energy of the interaction between the dressed bag and third nucleon; 
and
\item[--] quite significant correction to the kinetic energy of the system due to the $np$ mass 
difference and
high average kinetic energy.
\end{itemize}

The second general point that must be emphasized is a rather large admixture
of dibaryon-nucleon  components in both $^3$H and $^3$He, which has been
calculated in a completely consistent way. Closely  associated with the above
$6qN$ components, it is a specific energy dependence of the two-body force  in
a three- (and many-) body system. This energy dependence strongly reduces the
contribution of two- body force when a strong attractive three-body force is
added to the system Hamiltonian. This is a  manifestation of a very specific
new interplay between two- and three-body forces: the stronger the  {\em
three-body} force, the smaller the total contribution of the {\em two-body}
force to the nuclear  binding energy! By this way, a very natural density
dependence of nuclear interactions appears from  the beginning. Thus, the
general properties of the $3N$ system, where forces so much differ from any
conventional  model force, should appear also much differ from the predictions
of any conventional model and, hence, from  experiment.

It was very surprising to find that the characteristics of the $3N$ system in
our case turned out to be very  close to the predictions of the modern force
model (such as AV18 + UIX) and thus to experiment. This  gives us a good
test of the self-consistency and accuracy of the new force model. However,
predictions of the present $2N$- and $3N$-force model in other aspects will
strongly deviate  from those for conventional models. First, these are the
properties determined by the high-momentum  component of nuclear
wavefunctions. The point is that the system described by our multi-component 
wavefunctions explicitly including dibaryon components can easily absorb quite
high momentum  transfers, which can hardly be absorbed by the system described
by traditional multi-nucleon  wavefunctions. Therefore, to fit the
experimental data corresponding to large momentum transfers ($\sim\! 1\mbox{
GeV}/c$), many  types of meson-exchange and isobar currents are often
introduced to theoretical frameworks. However,  these currents are often
unrelated to the underlying force model. Hence, it is rather difficult to 
check the self-consistency of such calculations, e.g., the validity of gauge
invariance etc. 

Thus, the alternative description given here by the new force model can be
more self-consistent and  straightforward. One aspect of this new picture is
evident -- the present model applied to any  electromagnetic process on nuclei
automatically leads to a consistent whole picture of the process: single-nucleon 
currents at low momentum transfers, meson-exchange currents
(including new meson currents)  at intermediate momentum transfers, and quark
counting rules at very high momentum transfers, because  the model
wavefunction explicitly includes multinucleon, meson-exchange and multiquark
components.

%\bigskip
%\centerline{\bf Acknowledgments}

\ack
We are deeply grateful to numerous our colleagues for the fruitful discussions
and great assistance in  the calculations. We thank Drs. V.T. Voronchev, I.T.
Obukhovsky, Yu.M.Tchuvilsky and other members of  the Institute of Nuclear
Physics, Moscow State University, and Prof. P.Grabmayr and Dr. M.Kaskulov 
(Physikalisches Institut f\"ur Universit\"at T\"ubingen, T\"ubingen University). This work was supported in part
by der Deutsche  Forschungsgemeinschaft (grant No. Fa-67/20-1) and the Russian
Foundation for Basic Research (grant  Nos. 01-02-04015 and 02-02-16612).

\end{document}